\documentstyle[aps,prb]{revtex}
\begin{document}

\input epsf.sty
\twocolumn[\hsize\textwidth\columnwidth\hsize\csname %
@twocolumnfalse\endcsname

\draft

\widetext

\title{Magnetic field effects and magnetic anisotropy in lightly doped
La$_{2-x}$Sr$_x$CuO$_4$}
\author{M. Matsuda}
\address{
Advanced Science Research Center,
Japan Atomic Energy Research Institute, Tokai, Ibaraki 319-1195, Japan}
\author{M. Fujita and K. Yamada}
\address{
Institute for Chemical Research, Kyoto University, Gokasho, Uji
610-0011, Japan}
\author{R. J. Birgeneau}
\address{
Department of Physics, University of Toronto, Toronto, Ontario, Canada M5S 1A1}
\author{Y. Endoh}
\address{
Institute for Materials Research, Tohoku University, Katahira,
Sendai 980-8577, Japan}
\author{G. Shirane}
\address{
Department of Physics, Brookhaven National Laboratory, Upton, New York 11973
}

\date{\today}
\maketitle
\begin{abstract}
The effects of the application of a magnetic field on the diagonal
stripe spin-glass phase is studied in
lightly doped La$_{2-x}$Sr$_x$CuO$_4$ ($x$=0.014 and 0.024).
With increasing magnetic field,
the magnetic elastic intensity at the diagonal incommensurate (DIC) positions
(1,$\pm\epsilon$,0) decreases as opposed to the increase seen in
superconducting samples. This diminution in intensity with increasing
magnetic field originates from
a spin reorientation transition,
which is driven by the antisymmetric exchange term
in the spin Hamiltonian.
On the other hand, the transition temperature,
the incommensurability, and the peak width of the diagonal
incommensurate correlations are not changed with magnetic field.
This result suggests that the magnetic correlations are determined
primarily by the charge disproportionation and that the geometry of 
the diagonal incommensurate magnetism is also determined by effects,
that is, stripe formation which are not purely magnetic in origin.
The Dzyaloshinskii-Moriya antisymmetric exchange
is nevertheless important in determining
the local spin structure in the DIC stripe phase.

\end{abstract}

\phantom{.}
]
\narrowtext

\section{Introduction}
Extensive elastic neutron scattering studies have been performed
on lightly doped La$_{2-x}$Sr$_x$CuO$_4$ (0$\le x<$0.055)
in order to elucidate the static magnetic properties in the spin-glass
regime. The pioneering studies of Wakimoto $et$ $al.$
reveal that the static spin correlations in the spin-glass phase
show a one-dimensional diagonal spin modulation, in which the direction of
the modulation is rotated away by 45$^\circ$ from that in
the superconducting phase.~\cite{wakimoto0,wakimoto1}
In the lightly doped regime 0$\le x<$0.02, it is well established
that a three-dimensional (3D) antiferromagnetic (AF)
long-range ordered phase and a spin-glass phase coexist at low
temperatures.
Matsuda $et$ $al.$ suggested that in this regime
electronic phase separation of the doped holes occurs
so that some regions with hole concentration
$c\rm_h\sim$0.02 exhibit diagonal stripe correlations
while the rest with $c\rm_h\sim$0 shows 3D AF order.~\cite{matsuda2}

One of the remaining puzzles is that the nuclear superlattice
peaks, which are predicted by the charge stripe model
and are prominent in doped La$_2$NiO$_4$,~\cite{tranquada3}
have not been observed in the spin-glass phase of La$_{2-x}$Sr$_x$CuO$_4$
although diagonal incommensurate (DIC) magnetic peaks are readily
observed.
It is, of course, possible that the charge and magnetic stripe model
is not correct.
However, there are at least two possibilities to understand
the absence of observable nuclear superlattice peaks in La$_{2-x}$Sr$_x$CuO$_4$.
One is that the peak width is very broad because the charge stripes are
significantly disordered. Another is that the peak intensity is
extremely low because the coupling between the charge and the lattice
is weak.
Magnetic field experiments should give valuable information
about the origin of the incommensurate magnetic peaks.
If the incommensurate peaks are not directly related to underlying
charge stripes but rather are purely magnetic in origin, then application of
a magnetic field could directly affect the static magnetic order.
Magnetic field experiments are also important to understand
the role of the magnetic anisotropy as in pure La$_2$CuO$_4$.

It is well known that in pure La$_2$CuO$_4$
the net magnetic anisotropy is weakly Ising-like with a strong
XY anisotropy and a weaker in-plane Ising anisotropy with
the easy-axis along the $b$ axis.~\cite{peters}
The easy-plane anisotropy in the CuO$_2$ plane is due to the
combination of the spin-orbit
and Coulomb exchange interactions.~\cite{wohlman0,wohlman}
The Dzyaloshinskii-Moriya (D-M) antisymmetric exchange
terms are generated by the small rotation of the CuO$_6$ octahedra,
which creates the magnetic
anisotropy in the CuO$_2$ plane as well as the canting of the
Cu$^{2+}$ moments out of the plane.~\cite{thio,kastner}
In pure La$_2$CuO$_4$ application of a magnetic field perpendicular
to the CuO$_2$ plane causes a first order magnetic transition and,
in addition, a weak ferromagnetic component is
induced along the magnetic field direction.
The magnetic structure also changes
from an La$_2$CuO$_4$-type (propagation vector
$\mbox{\boldmath $\hat{\tau}$}$=$\mbox{\boldmath $\hat{a}$}$) to an
La$_2$NiO$_4$-type
($\mbox{\boldmath $\hat{\tau}$}$=$\mbox{\boldmath $\hat{b}$}$)
in the high field phase.~\cite{kastner}

With hole doping, the magnetic anisotropy is expected to become
smaller since the rotation of the CuO$_6$ octahedra is reduced.
It was previously believed that when the long-range magnetic order
disappears at high temperatures
or at $x>$0.02, the magnetic anisotropy plays a
less significant role in the magnetic properties.
It has been reported, however, that the easy-plane anisotropy
is required to describe the $L$-dependence of the neutron elastic
magnetic intensity,~\cite{matsuda0,matsuda1}
in La$_{2-x}$Sr$_x$CuO$_4$ for $x\ge$0.02.
It has also been found from magnetization measurements using untwinned
crystals that even the easy-axis anisotropy is evident above
$T\rm_{N}$ in crystals with $x$=0, 0.01, and 0.02 and at all
temperatures in an $x$=0.03 sample.~\cite{lavrov}

In this study we have examined the effect of a magnetic field on
the static magnetic
correlations in lightly doped La$_{2-x}$Sr$_x$CuO$_4$.
One important point we wished to clarify is whether the DIC
structure is dominated by charge disproportionation or is purely
magnetic in origin.
It is also important to clarify the nature of the magnetic anisotropy
in doped La$_2$CuO$_4$ from a microscopic point of view.
Although the DIC structure is probably dominated by
the charge disproportionation, the static magnetic ordering is in part
stabilized by the magnetic anisotropy.

\section{Experimental Details}
The single crystals of La$_{2-x}$Sr$_x$CuO$_4$ ($x$=0.014 and
0.024) were grown by the traveling solvent floating-zone method.
The crystals were annealed in an Ar atmosphere at 900 $^\circ$C for 24 h.
The uncertainty in the effective hole
concentration of each crystal was estimated to be less than 10\%.
The crystals used in this study were the ones that were employed
in the previous neutron-scattering studies.~\cite{matsuda2,matsuda1}
The twin structure of the crystals is shown in Fig. 1.
As shown in Fig. 2 of Ref. \onlinecite{wakimoto1}, four twins are
possible in the low temperature orthorhombic phase ($Bmab$).
A smaller number of twins greatly simplifies the analysis of
the incommensurate
magnetic peak structure. Fortunately, the $x$=0.014 crystal
has a single domain dominant. 
The $x$=0.024 crystal has two twins, which are estimated to be equally
distributed based on the ratio of the nuclear Bragg peak intensities
from each twin.
\begin{figure}
\centerline{\epsfxsize=3.3in\epsfbox{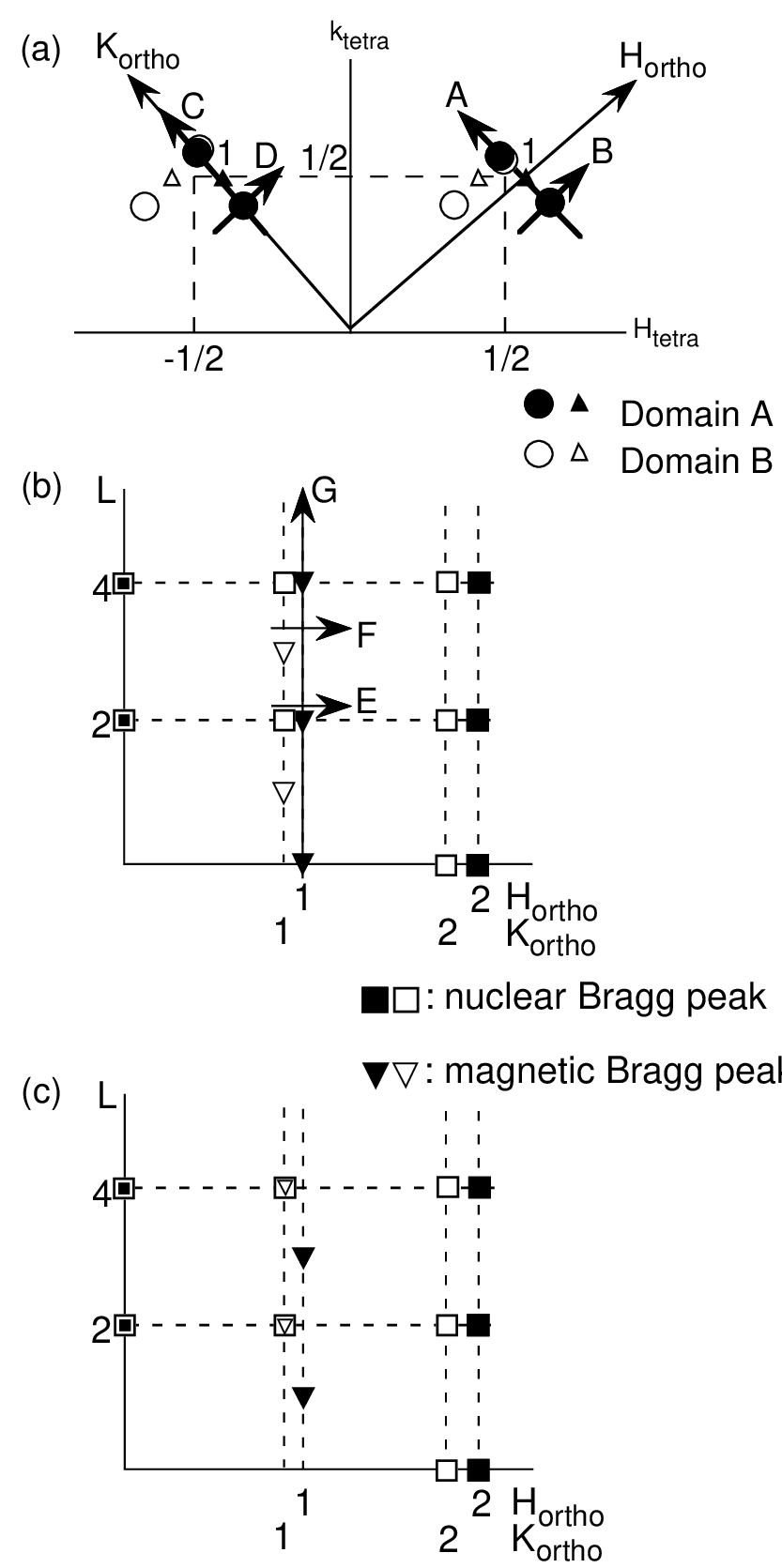}}
\caption{(a) Diagram of the reciprocal lattice in the $(HK0)$
scattering zone. Filled and open symbols are for domains A and B,
respectively. The circles and triangles correspond to the incommensurate
magnetic peaks and fundamental Bragg peaks, respectively.
(b) Diagram of the reciprocal lattice in the $(H0L)$
scattering zone. The magnetic reflections predicted from the
La$_2$CuO$_4$-type structure are shown.
Filled and open symbols are for domains A and B,
respectively. The triangles and squares correspond to the
magnetic peaks and nuclear Bragg peaks, respectively.
(c) The magnetic reflections predicted
from the La$_2$NiO$_4$-type structure are shown.}
\label{fig1}
\end{figure}

The neutron-scattering experiments were carried out on the
cold neutron three-axis spectrometer LTAS and the thermal neutron
three-axis spectrometer TAS2 installed in the guide hall of JRR-3M at
the Japan Atomic Energy Research Institute.
Typical horizontal collimator sequences were
guide-80$'$-S-20$'$-80$'$ with a fixed incident neutron energy of
$E\rm_i$=5.05 meV at LTAS and guide-40$'$-S-40$'$-80$'$ with a fixed
incident neutron energy of $E\rm_i$=13.7 meV at TAS2.
Contamination from higher-order beams was effectively eliminated
using Be filters at LTAS and pyrolytic graphite filters at TAS2.
The single crystals were oriented in the $(HK0)\rm_{ortho}$ or
$(H0L)\rm_{ortho}$ scattering plane.
The neutron-scattering experiments in a magnetic field were
performed up to 10 T using a new type of split-pair superconducting 
magnet cooled by cryocoolers.
The field was applied perpendicular to the 
scattering plane.
In this paper, we use the low temperature
orthorhombic phase ($Bmab$) notation $(h,k,l)\rm_{ortho}$
to express Miller indices.

\section{Results}
\subsection{Magnetic field perpendicular to the CuO$_2$ plane}
\subsubsection{La$_{1.986}$Sr$_{0.014}$CuO$_4$
(N\'{e}el ordered phase)}
The effect of a magnetic field was studied first
in La$_{1.986}$Sr$_{0.014}$CuO$_4$,
which exhibits coexistence of a 3D N\'{e}el ordered phase and
a quasi-2D spin-glass phase
at low temperatures.~\cite{matsuda2}
Figure 2 shows the magnetic field dependence of
the magnetic Bragg intensity at (1,0,0) in
La$_{1.986}$Sr$_{0.014}$CuO$_4$. The magnetic field is applied
perpendicular to the CuO$_2$ plane.
At 50 K the (1,0,0) magnetic Bragg intensity, originating from the
long-range AF ordering, gradually decreases with
increasing magnetic field and shows a sharp drop at
a critical field of 4.4 T.
The magnetic structure factor of the (1,0,0) reflection becomes zero
because a spin reorientation transition from the La$_2$CuO$_4$-type 
magnetic structure to
the La$_2$NiO$_4$-type magnetic structure occurs.~\cite{kastner}
In this transition the spins in the orthorhombic unit cell at (0,0,0)
($\mbox{\boldmath $\rm{S}$}$$\parallel$$\mbox{\boldmath $\hat{b}$}$)
and ($\frac{1}{2}$,$\frac{1}{2}$,0)
($\parallel$-$\mbox{\boldmath $\hat{b}$}$) are unchanged at $H_c$ but those
at (0,$\frac{1}{2}$,$\frac{1}{2}$) ($\parallel$$\mbox{\boldmath $\hat{b}$}$)
and
($\frac{1}{2}$,0,$\frac{1}{2}$) ($\parallel$-$\mbox{\boldmath $\hat{b}$}$)
for $H<H_c$ change sign above $H_c$.
This critical field is comparable to that in La$_2$CuO$_{4+\delta}$,
which also has a similar N\'{e}el temperature $T_N$=234 K.~\cite{kastner}

With decreasing temperature, the critical field becomes larger.
At the same time, magnetic hysteresis occurs between the field
increasing and field decreasing processes as shown in Fig. 2(a).
We confirmed that this behavior is reproducible.
A characteristic feature is that the intensity at (1,0,0) does not return
to the initial value even at 0 T after the magnetization cycle.
Such large hysteresis behavior at (1,0,0) was not observed
in La$_2$CuO$_{4+\delta}$.~\cite{kastner}
A possible origin will be discussed in Sec. 4.
\begin{figure}
\centerline{\epsfxsize=3.2in\epsfbox{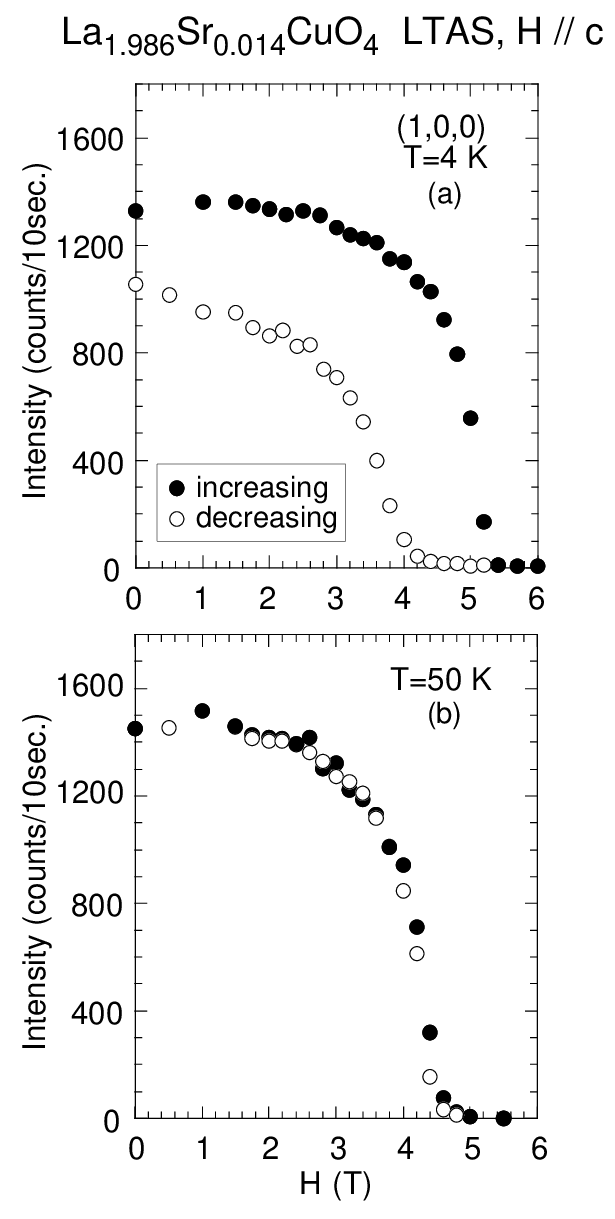}}
\caption{Magnetic field dependence of the (1,0,0)
magnetic Bragg intensity at 4 K and 50 K
in La$_{1.986}$Sr$_{0.014}$CuO$_4$. Magnetic field is applied
perpendicular to the CuO$_2$ plane.}
\label{fig2}
\end{figure}

\twocolumn[\hsize\textwidth\columnwidth\hsize\csname %
@twocolumnfalse\endcsname
\widetext
\begin{figure*}
\centerline{\epsfxsize=7in\epsfbox{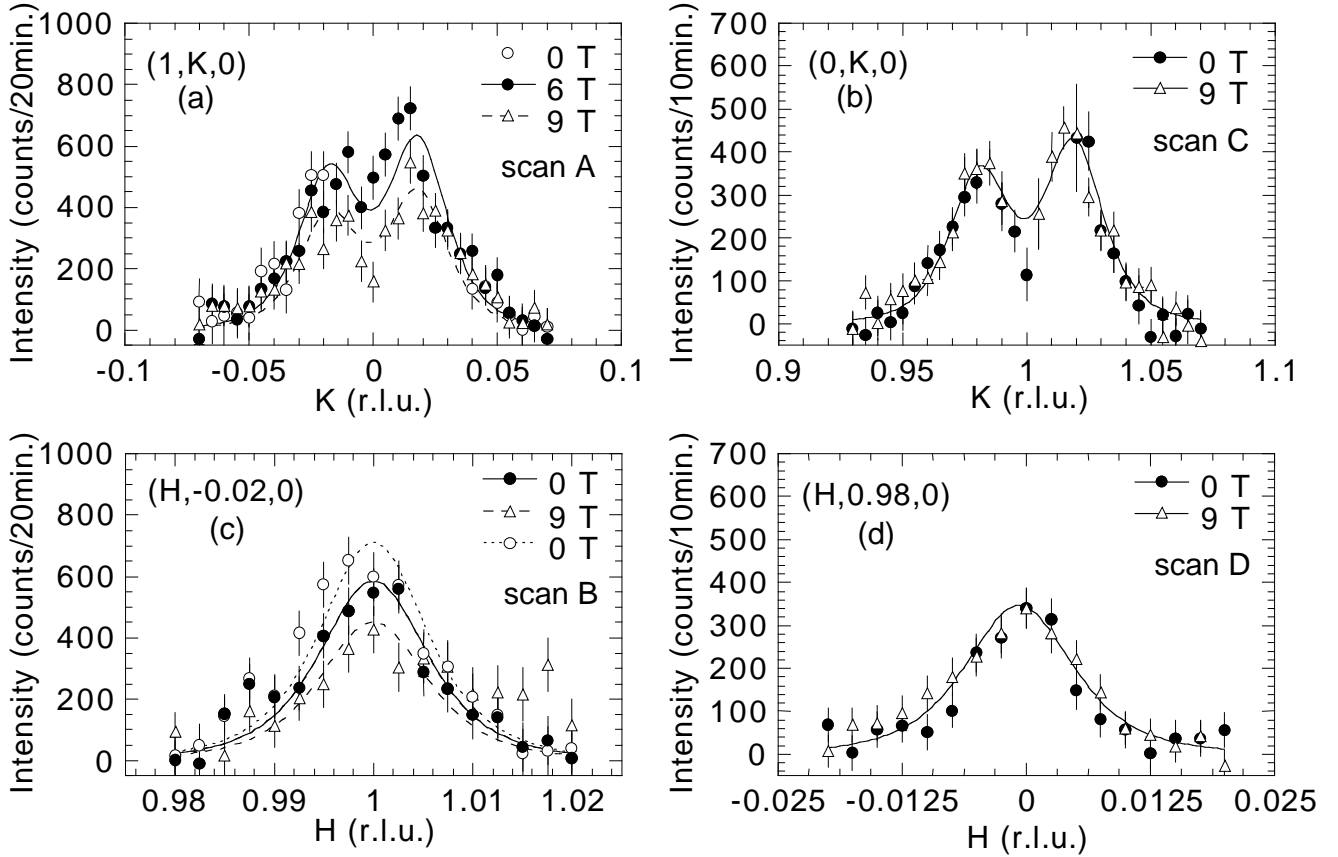}}
\caption{Magnetic elastic intensity around the diagonal
incommensurate positions (1,$\pm\epsilon$,0) and (0,1$\pm\epsilon$,0)
at 4 K in La$_{1.986}$Sr$_{0.014}$CuO$_4$ as a function of magnetic field.
The scan trajectories are shown in Fig. 1(a).
Open circles in (c) are the data measured at 0 T after a magnetization
cycle. Magnetic field is applied perpendicular to the CuO$_2$ plane.
Background intensities measured at a high temperature have been
subtracted. The lines are guides to the eyes.}
\label{fig3}
\end{figure*}
\phantom{.}
]
\narrowtext
\subsubsection{La$_{1.986}$Sr$_{0.014}$CuO$_4$
(spin-glass phase)}
Figure 3 shows the magnetic field dependence of the magnetic elastic
intensity around the DIC positions (1,$\pm\epsilon$,0)
and (0,1$\pm\epsilon$,0) in La$_{1.986}$Sr$_{0.014}$CuO$_4$.
The magnetic signal, which develops below $\sim$30 K, originates
from the phase-separated spin-glass regions.
The incommensurate peaks are asymmetric because of the twin
structure as shown in Fig. 1(a). The peak profiles are consistent with
those observed in the previous measurements at zero field.~\cite{matsuda2}
Below $H_c$ there 
exists an intense magnetic Bragg peak at (1,0,0) so that the data around
(1,0,0) are contaminated. On the other hand, above $H_c$, the data
around (0,1,0) are contaminated because of double scattering
at the (0,1,0) magnetic peak position.
This double scattering is made allowed by the transition to
the La$_2$NiO$_4$-type magnetic structure at high field.
As shown in Fig. 3(a), the magnetic intensities
at (1,$\pm\epsilon$,0)
are almost constant between 0 and 6 T and gradually decrease above 6 T.
Since the (1,0,0) peak in the N\'{e}el ordered phase shows
a decrease in intensity with increasing field,
it is natural to think that the decrease of the DIC peak intensity
also originates from the spin reorientation transition.
As shown in Fig. 2, $H_c$=5.2 T at 4 K in the N\'{e}el ordered phase,
indicating that the averaged transition field is similar, albeit somewhat
larger, in the spin-glass phase.
The transition is very broad in the spin-glass phase
because the cluster size is finite and, in particular, there is only
short range magnetic order ($\sim$10 \AA) between the CuO$_2$ planes.
This result is consistent with that of magnetization measurements,
which shows a broadening of the jump
in the magnetization at $H_c$.~\cite{suzuki}

The magnetic intensity at (0,1$\pm\epsilon$,0) is almost
magnetic field independent as shown in Figs. 3(b) and 3(d).
The magnetic signal at (0,1$\pm\epsilon$,0) may be considered to be
the tail of (0,1$\pm\epsilon$,1), which is broadened along
the $c$ axis because of the quasi-2D nature of
the DIC state.~\cite{matsuda1}
The instrumental resolution which is elongated vertically
integrates the tail effectively.
The (0,1,1) magnetic Bragg peak should decrease in intensity under magnetic
field, whereas the (0,1,0) magnetic intensity increases.
Both contributions will tend to compensate each other, so that
the intensity at (0,1$\pm\epsilon$,0) may remain almost unchanged.
\begin{figure}
\centerline{\epsfxsize=3.2in\epsfbox{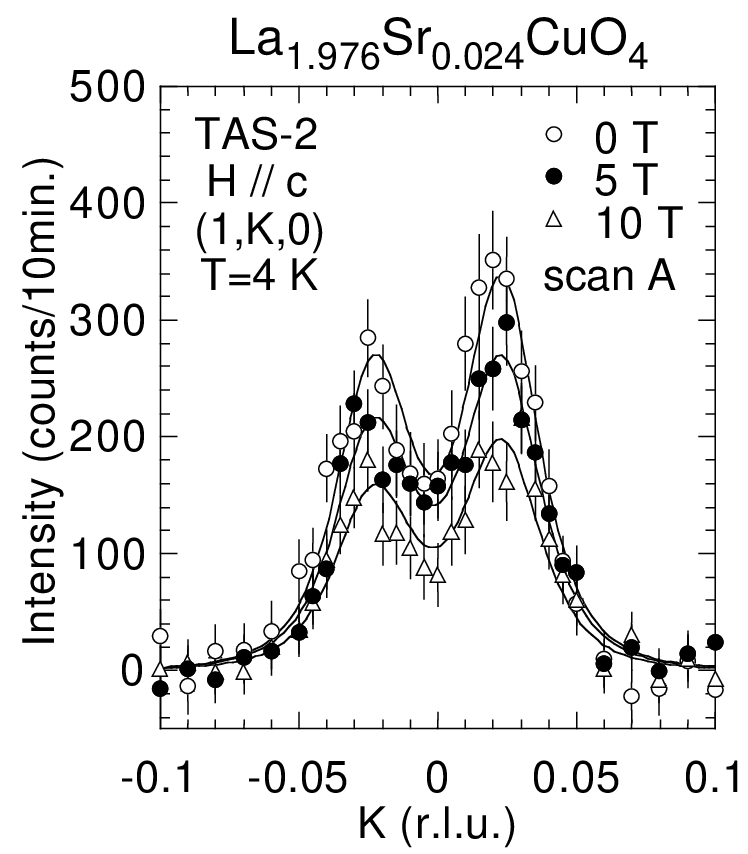}}
\caption{Magnetic elastic intensity around the diagonal
incommensurate positions (1,$\pm\epsilon$,0) at 4 K
in La$_{1.976}$Sr$_{0.024}$CuO$_4$ as a function of magnetic field.
The scan trajectory is shown in Fig. 1(a).
Magnetic field is applied perpendicular to the CuO$_2$ plane.
Background intensities measured at a high temperature have been
subtracted. The solid lines are guides to the eyes.}
\label{fig4}
\end{figure}

\subsubsection{La$_{1.976}$Sr$_{0.024}$CuO$_4$ (spin-glass phase)}
The effect of a magnetic field has also been studied
in La$_{1.976}$Sr$_{0.024}$CuO$_4$. At zero field,
this sample shows no long-range AF order but only diagonal stripe
spin-glass behavior
below $\sim$25 K. Figure 4 shows the effect of a magnetic field in
La$_{1.976}$Sr$_{0.024}$CuO$_4$ with the field perpendicular to
the CuO$_2$ plane. The magnetic intensity at (1,$\pm\epsilon$,0)
decreases monotonically with increasing magnetic field. The tendency
is similar to that in La$_{1.986}$Sr$_{0.014}$CuO$_4$
although the overall effect is much more gradual
presumably because the between-plane magnetic correlation length
is only 3 \AA.
Since this sample shows only spin-glass ordering at low temperatures,
the (1,0,0) magnetic Bragg peak is absent and the DIC
peaks per unit sample volume are more intense than those
in the $x$=0.014 sample.
Therefore, the magnetic field dependence of the DIC peaks can be
cleanly observed in this sample.
We found, in addition, that the transition temperature, the peak width,
and peak positions are almost unchanged under magnetic field.
\begin{figure}
\centerline{\epsfxsize=3.2in\epsfbox{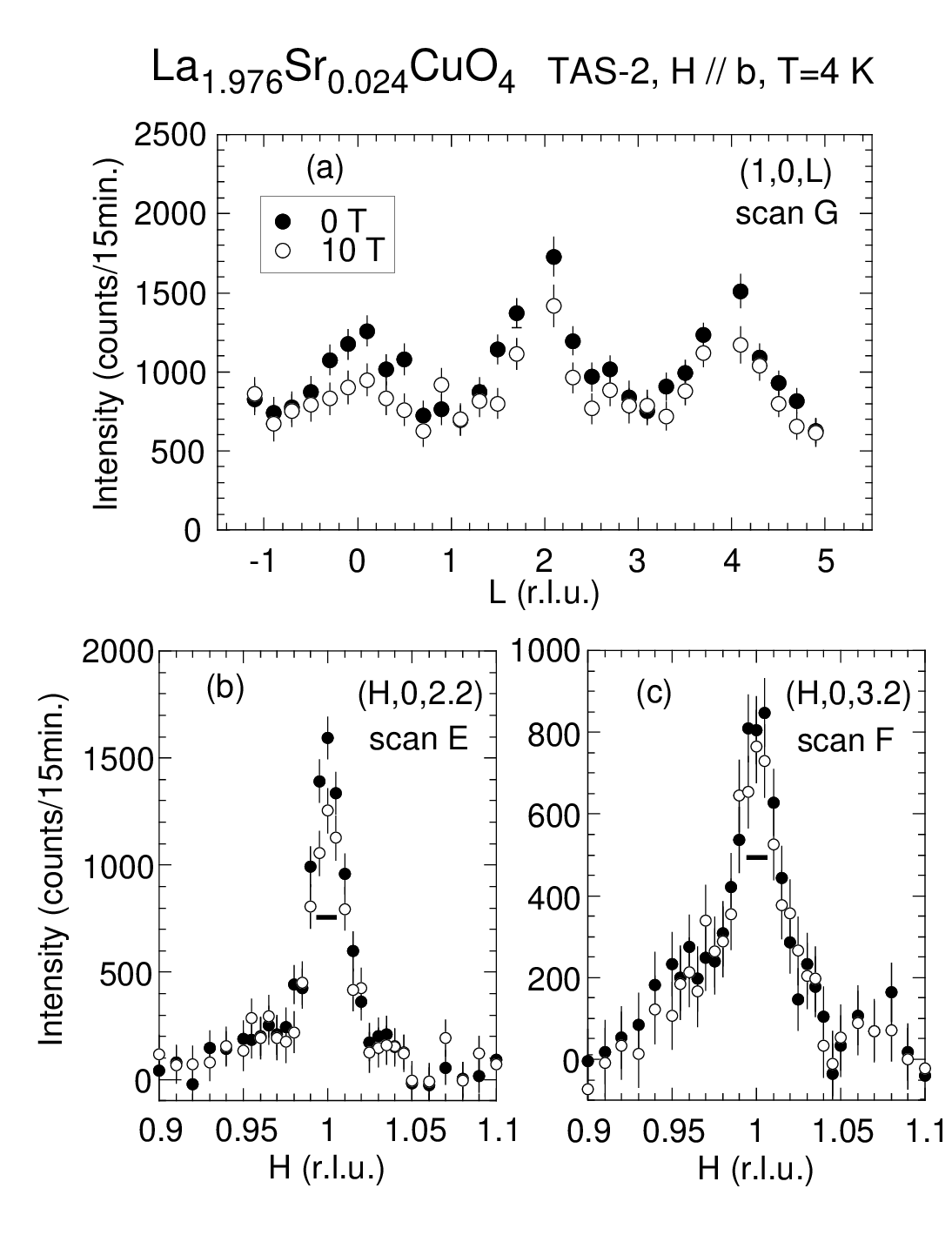}}
\caption{Magnetic elastic intensity at (1,0,$L$), ($H$,0,2.2),
and ($H$,0,3.2) at 4 K in La$_{1.976}$Sr$_{0.024}$CuO$_4$
as a function of magnetic field.
The scan trajectories are shown in Fig. 1(b).
Magnetic field is parallel to the CuO$_2$ plane.
Background intensities measured at a high temperature have been
subtracted.
The horizontal bars in (b) and (c) represent
the instrumental resolution.}
\label{fig5}
\end{figure}

\subsection{Magnetic field parallel to the CuO$_2$ plane}
We have so far discussed only the effects of a magnetic field
when the field is applied perpendicular to the CuO$_2$ plane.
Figure 5 shows the effect of the magnetic field applied parallel to
the CuO$_2$ plane in La$_{1.976}$Sr$_{0.024}$CuO$_4$.
The scan along (1,0,$L$) shows broad peaks at the $L=even$ positions
in zero field;~\cite{matsuda0}
these are the positions where magnetic Bragg peaks are observed
in pure La$_2$CuO$_4$.
The peaks are broad because the correlation length along the
$c$ axis is very short as noted above.
The magnetic peaks in Figs. 5(b) and 5(c) are asymmetric
and have a tail at lower $Q$'s because of the twin
structure as shown in Fig. 1(b). The peak profiles are consistent with
those observed in the previous measurements at zero field.~\cite{matsuda0}

Under a magnetic field of 10 T, the peak widths along the
$a$, $b$, and $c$ axes are unchanged, indicating that magnetic
correlation length is not measurably affected.
On the other hand, the magnetic intensity decreases
over a wide range of $L$.
In La$_2$CuO$_{4+\delta}$, a spin-flop transition, which is due to 
the D-M exchange, occurs when
the magnetic field is applied along the easy-axis $b$.~\cite{thio2}
Above the transition field, which is about twice as large as that
observed for
$\mbox{\boldmath $H$}$$\parallel$$\mbox{\boldmath $\hat{c}$}$,
spin components along the $a$ and $c$ axes appear.
This explains the decrease in magnetic intensity at (1,0,$L$)
qualitatively.

Thus, the magnetic elastic signal decreases under magnetic field irrespective
of the field direction. This result with $x$=0.024 sample
is consistent with that expected
from a broadened spin reorientation transition, originating from
the D-M exchange interaction, in analogy to the first order transition
in La$_2$CuO$_{4+\delta}$. Again, the transition is severely
broadened because of the very short between-plane correlation length.

\section{Discussion and Conclusions}
Our elastic neutron-scattering study in lightly doped
La$_{2-x}$Sr$_x$CuO$_4$ reveals
that the magnetic field affects the direction which
the spins point along. On the other hand, the transition temperature,
the peak width, and the
incommensurability are almost unaffected. This result suggests
that the magnetic correlations are dominated by the charge
disproportionation rather than magnetic interaction effects.

We have found that the DIC
structure in the spin-glass phase has an in-plane Ising anisotropy
originating from a combination of XY anisotropy and the D-M
antisymmetric exchange as in pure La$_2$CuO$_4$.
This result is consistent with that in the magnetization
measurements,~\cite{lavrov} which suggests that the in-plane magnetic
anisotropy continues to play an important role even
after the N\'{e}el order is destroyed.
It should be emphasized that our study evinced the in-plane
magnetic anisotropy specifically in the DIC phase.
This magnetic anisotropy plays an important role in determining
the spin direction in the DIC structure.
Since the rotation of the CuO$_6$ octahedra is larger in the lower
hole concentration region, the D-M antisymmetric exchange
will be more pronounced in the lower hole concentration region.
It has been reported previously that the static DIC spin modulation
is stabilized in the lower hole concentration region.~\cite{matsuda2}
This is partly because the diagonal stripe is increasingly stabilized
as the rotation of the CuO$_6$ octahedra becomes larger in the
$Bmab$ phase.
However, since the magnetic anisotropy, which reduces
quantum spin fluctuations, enhances hole attraction and
stabilizes magnetic order,~\cite{riera,chernyshev,riera2}
the static DIC structure in
La$_{2-x}$Sr$_x$CuO$_4$ may also be influenced by the magnetic anisotropy.
It would be interesting to observe directly
the energy gap in the spin excitations due to the magnetic anisotropy
at the magnetic zone center. However, the peak width
in energy becomes so broad that the anisotropy
gap energy is very difficult to determine experimentally.~\cite{matsuda1}

The magnetic anisotropy in lightly doped La$_{2-x}$Sr$_x$CuO$_4$
should be taken into account in determining the appropriate spin structure
model in the spin-glass phase.
The $L$-dependence of the magnetic intensity, shown in Fig. 5(a),
can be described by 3D correlated AF spin clusters with the spin
randomly oriented within the $ab$ plane and with
$\mbox{\boldmath $\hat{\tau}$}$($\parallel$$\mbox{\boldmath $\hat{a}$}$).
Alternatively, the data are consistent with equal admixtures of
3D correlated phases where the spin vector
$\mbox{\boldmath $\rm{S}$}$ is along or perpendicular to
$\mbox{\boldmath $\hat{\tau}$}$($\parallel$$\mbox{\boldmath $\hat{a}$}$)
.~\cite{matsuda0}
It should be noted that all the spins point along the $b$ axis
in excess-oxygen-doped La$_2$CuO$_{4+y}$,
which shows a parallel incommensurate magnetic
modulation below $\sim$42 K.~\cite{lee0}

Lee $et$ $al.$ suggested that the weak ferromagnetic moment in each
CuO$_2$ plane, originating from the out-of-plane spin canting,
is absent in La$_2$CuO$_{4+y}$
because in the stripe model, in which there exist antiphase domain boundaries
between ordered spins, the out-of-plane spin canting points in opposite
directions on either side of a domain wall.~\cite{lee0}
Then, the size of the magnetic field effect would be inversely related to
the number of antiphase domains inside of a magnetically correlated area,
suggesting that the decrease in magnetic intensity in a magnetic field
should become smaller as the Sr-doping gets larger.
There may, of course, be additional magnetic mechanisms,
in addition to the one we have identified, contributing to
the diminution of the DIC magnetic peak intensity under the
application of a magnetic field.

The effects of a magnetic field have also been studied in superconducting
La$_{2-x}$Sr$_x$CuO$_4$ ($x$=0.10 and 0.12) and
excess-oxygen-doped La$_2$CuO$_{4+y}$.
It has been found that the static parallel
stripe magnetic order is enhanced under the application of a magnetic field
perpendicular to the CuO$_2$ planes.~\cite{katano,lake,lee}
This is completely opposite to our result in lightly doped
La$_{2-x}$Sr$_x$CuO$_4$. The enhancement of the elastic magnetic
intensity is ascribed to the vortices which stabilize the static
magnetic order over a region much larger than
the vortex cores.~\cite{katano,lake,lee}
Our results indirectly support this interpretation.

We now discuss the magnetic hysteresis behavior in
La$_{1.986}$Sr$_{0.014}$CuO$_4$. The hysteresis behavior
becomes pronounced with decreasing temperature below 50 K.
This can be related to the spin-glass behavior.
As shown in Fig. 3(c), the magnetic intensity at (1,$-$0.02,0)
increases slightly after a magnetization cycle.
The slight increase in magnetic intensity after the magnetization cycle
is probably related to the (1,0,0) magnetic Bragg intensity
lost after the cycle, as shown in Fig. 2(a). However, the increased intensity
at (1,$-$0.02,0) is factor of $\sim$30 smaller than
the lost intensity at (1,0,0).
Therefore, some fraction of the N\'{e}el ordered phase still remains in
the La$_2$NiO$_4$-type structure after the magnetization cycle.

In summary, our neutron-scattering experiments under magnetic field
demonstrate the effects of the D-M interaction in the spin-glass
phase in lightly doped La$_{2-x}$Sr$_x$CuO$_4$.
Although the DIC structure regimes are determined by
the charge disproportionation, the static magnetic ordering is
stabilized in part by the magnetic anisotropy.

\section*{Acknowledgments}
We would like to thank S. Katano, Y. S. Lee, T. Suzuki, and S. Wakimoto
for stimulating discussions and Y. Shimojo for technical assistance.
This study was supported in part by the
U.S.-Japan Cooperative Program on Neutron Scattering, by a Grant-in-Aid
for Scientific Research from the Japanese Ministry of Education, Science,
Sports and Culture, by a grant for the promotion of science from the
Science and Technology Agency, and by CREST.
Work at Brookhaven National Laboratory was carried out under Contract
No. DE-AC02-98CH10886, Division of Material Science, U.S. Department of
Energy.
Work at the University of Toronto is part of the Canadian
Institute for Advanced Research and is supported by the Natural Science
and Engineering Research Council of Canada.

\end{document}